%% file: main.tex
\documentclass[runningheads]{llncs}

\input{./header.tex}

\begin{document}
\title{Leveraging \tla\ Specifications to Improve\\ the Reliability of the \zk \\ Coordination Service 
\thanks{
This work is supported by the National Natural Science Foundation of China (62372222), the CCF-Huawei Populus Grove Fund (CCF-HuaweiFM202304), the Cooperation Fund of Huawei-Nanjing University Next Generation Programming Innovation Lab (YBN2019105178SW38), the Fundamental Research Funds for the Central Universities (020214912222) and the Collaborative Innovation Center of Novel Software Technology and Industrialization. 
Yu Huang is the corresponding author.
}
}

\titlerunning{Leveraging \tla\ to Improve \zk\ Reliability}
%

\author{Lingzhi Ouyang \orcidID{0000-0001-7523-8759} 
\and Yu Huang \orcidID{0000-0001-8921-036X} 
\and \\ Binyu Huang \orcidID{0009-0007-8359-9010}
\and Xiaoxing Ma \orcidID{0000-0001-7970-1384}
}

\authorrunning{L. Ouyang et al.}
%
\institute{State Key Laboratory for Novel Software Technology, Nanjing 210023, China
\email{\{lingzhi.ouyang, binyuhuang\}@smail.nju.edu.cn, \{yuhuang,xxm\}@nju.edu.cn}\\
}
\maketitle              

\input{./sections/0_abs.tex}


\input{./sections/1_intro.tex}

\input{./sections/2_preli.tex}

\input{./sections/3_prot-spec.tex}

\input{./sections/4_syst-spec.tex}

\input{./sections/5_test-spec.tex}

\input{./sections/6_related-work.tex}

\input{./sections/7_concl.tex}


\bibliographystyle{splncs04}
\bibliography{main}

\end{document}

%% file: header.tex
\usepackage[T1]{fontenc}

\usepackage{graphicx}
\usepackage[svgnames]{xcolor}
\usepackage{CJK}
\usepackage[colorlinks,
            linkcolor=blue,
            anchorcolor=blue,
            citecolor=blue]{hyperref}

\usepackage{booktabs}
\usepackage{makecell}
\usepackage{amssymb}

\usepackage{multirow}
\usepackage{threeparttable}

\newcommand{\zk}{ZooKeeper}
\newcommand{\tla}{TLA$^+$}
\newcommand{\met}{\textsc{Met}}

\newcommand{\bla}{\color{black}}

%% file: sections/0_abs.tex
\begin{abstract}

\zk\ is a coordination service, widely used as a backbone of various distributed systems.
Though its reliability is of critical importance, testing is insufficient for an industrial-strength system of the size and complexity of \zk, and deep bugs can still be found.
To this end, we resort to formal \tla\ specifications to further improve the reliability of \zk.
Our primary objective is usability and automation, rather than full verification.
We incrementally develop three levels of specifications for \zk. %
We first obtain the \textit{protocol specification}, which unambiguously specifies the Zab protocol behind \zk.
We then proceed to a finer grain and obtain the \textit{system specification}, which serves as the super-doc for system development.
In order to further leverage the model-level specification to improve the reliability of the code-level implementation, we develop the \textit{test specification}, which guides the explorative testing of the \zk\ implementation.
The formal specifications help eliminate the ambiguities in the protocol design and provide comprehensive system documentation. They also 
help find critical deep bugs 
in system implementation, which are beyond the reach of state-of-the-art testing techniques.
Our specifications have been merged into the official Apache \zk\ project.

\keywords{\tla \and \zk\ \and Zab \and Specification \and Model checking} 

\end{abstract}

%% file: sections/1_intro.tex
\section{Introduction}

\zk\ is a distributed coordination service for highly reliable synchronization of cloud applications \cite{Hunt10}.
\zk\ essentially offers a hierarchical key-value store, which is used to provide a distributed configuration service, synchronization service, and naming registry for large distributed systems.
Its intended usage requires Zookeeper to provide strong consistency
guarantees, which it does by running a distributed consensus protocol called Zab \cite{Junqueira11}.

Consensus protocols are notoriously difficult to get right.
The complex failure recovery logic results in an astronomically large state space, and deep ``Heisenbugs" still often escape from intensive testing \cite{Lees14,Lees15}.
Toward this challenge, we resort to formal methods to improve the reliability of \zk.
We do not aim to achieve full verification, but instead emphasize a high degree of automation and practical usability.
Moreover, our primary goal is to improve the reliability of both the model-level design and the code-level implementation.

We adopt \tla\ as our specification language.
\tla\ has been successful in verifying distributed concurrent systems, especially consensus protocols.
Many consensus protocols, including Paxos, Raft and their derivatives, have their \tla\ specifications published along with the protocol design and implementation \cite{Paxos-TLA-v1,PaxosTLA2019,Raft-TLA,Gafni03}.
However, the current usage of \tla\ is mainly restricted to verification of the protocol design.
Considering code-level implementation, model checking-driven test case generation is used to ensure the equivalence between two different implementations in MongoDB Realm Sync \cite{Davis20vldb}.

Our primary objective is to improve the reliability of \zk.
We incrementally obtain three levels of specifications in \tla.
We first obtain the \textit{protocol specification}, which unambiguously specifies the Zab protocol behind \zk.
We then proceed to a finer grain and obtain the \textit{system specification}, which serves as the super-doc
\footnote{
Super-doc refers to the precise, concise and testable documentation of the system implementation, which can be explored and experimented on with tools
\cite{Newcombe15}.} 
for \zk\ development.
In order to leverage the model-level specification to improve the reliability of the code-level implementation, we further develop the \textit{test specification}, which guides the explorative testing of the \zk\ implementation.
The formal specifications help eliminate the ambiguities in the protocol design and provide comprehensive system documentation. They also 
help find critical deep bugs in system implementation, which are beyond the reach of state-of-the-art testing techniques.
Our specifications are available in the online repository \cite{ZK-TLA-Spec}.
Writing \tla\ specifications for \zk\ was raised as an issue \cite{ZK-Issue3615}.
Our specifications have addressed this issue and been accepted by the Apache \zk\ project \cite{apache-zk-spec}.
\bla

We start in Section \ref{Sec:Preli} by introducing the basics of \zk\ and \tla.
We present our three levels of specifications in Sections \ref{Sec:Prot-Spec}, \ref{Sec:Syst-Spec} and \ref{Sec:Test-Spec}.
In Section \ref{Sec:Related-work}, we discuss the related work. 
Finally, Section \ref{Sec:Concl} concludes with a discussion on the benefits of and the potential extensions to our formal specification practices.

%% file: sections/2_preli.tex
\section{\zk\ and \tla} \label{Sec:Preli}

\subsection{\zk\ and Zab}

\zk\ is a fault-tolerant distributed coordination service used by a variety of distributed systems \cite{ZK-homepage,Hunt10,Medeiros12}. 
These systems often consist of a large number of processes and rely upon \zk\ to perform essential coordination tasks, such as maintaining configuration information, storing status of running processes and group membership, providing distributed synchronization and managing failure recovery.

Due to the significant reliance of large applications on \zk, the service must maintain a high level of availability and possess the ability to mask and recover from failures.
\zk\ achieves availability and reliability through replication and utilizes a primary-backup scheme to maintain the states of replica processes consistent \cite{Junqueira11,Junqueira10,Zab-wiki}. 
Upon receiving client requests, the primary generates a sequence of non-commutative state changes and propagates them as transactions to the backup replicas using \textit{Zab}, the \zk\ atomic broadcast protocol. 
The protocol consists of three phases: \textsf{DISCOVERY}, \textsf{SYNC} and \textsf{BROADCAST}. 
Its primary duties include agreeing on a leader in the ensemble, synchronizing the replicas, managing the broadcast of transactions, and recovering from failures.

To ensure progress, \zk\ requires that a majority (or more generally a quorum) of processes have not crashed.
Any minority of processes may crash at any moment. 
Crashed processes are able to recover and rejoin the ensemble.
For a process to perform the primary role, it must have the support of a quorum of processes. 

\subsection{\tla\ Basics}

\tla\ (Temporal Logic of Actions) is a lightweight formal specification language that is particularly suitable for designing distributed and concurrent systems \cite{tla-website}.
In contrast to programming languages, \tla\ employs simple mathematics to express concepts in a more elegant and precise manner.
In \tla, a system is specified as a state machine by defining the possible initial \textit{states} and the allowed \textit{actions}, i.e., state transitions.
Each state represents a global state of the system.
Whenever all \textit{enabling conditions} of a state transition are satisfied in a given \textit{current} state, the system can transfer to the \textit{next} state by applying the action.

One of the notable advantages of \tla\ is its ability to handle different levels of abstraction. 
Correctness properties and system designs can be regarded as steps on a ladder of abstraction, with correctness properties occupying higher levels, system designs and algorithms in the middle, and executable code at the lower levels \cite{Newcombe15}.
The flexibility to choose and adjust levels of abstraction makes \tla\ a versatile tool suited to a wide range of needs. 

\tla\ provides the TLC model checker, which builds a finite state model from the specifications for checking invariant safety properties (in this work, we mainly focus on safety properties and do not consider liveness properties). 
TLC first generates a set of initial states that satisfy the specification, and then traverses all possible state transitions. 
If TLC discovers a state violating an invariant property, it halts and provides the trace leading to the violation.
Otherwise, the system is verified to satisfy the invariant property. 
With TLC, we are able to explore every possible behavior of our specifications without additional human effort.
As a result, we can identify subtle bugs and edge cases that may not be exposed through other testing or debugging techniques.

%% file: sections/3_prot-spec.tex
\section{Protocol Specification} \label{Sec:Prot-Spec}

We first develop the \textit{protocol specification}, i.e., specification of the Zab protocol based on its informal description \cite{Junqueira11,Hunt10}.
The protocol specification aims at precise description and automatic model checking of the protocol design.
It also serves as the basis for further refinements, as detailed in Sections \ref{Sec:Syst-Spec} and \ref{Sec:Test-Spec}.

\subsection{Development of Protocol Specification}
The protocol specification necessitates a comprehensive description of the design of Zab, along with the corresponding correctness conditions that must be upheld. In the following, we present our practice of developing the \tla\ specifications for both aspects.
\bla

\subsubsection{Specification of Zab.}
The ``upon-do clauses" in the Zab pseudocode can be readily transformed to the enabling conditions and actions in \tla.
This feature greatly simplifies the task of obtaining the initial skeleton of the protocol specification.
The real obstacle lies in handling the ambiguities and omissions in the informal design. We have three challenges to address, as detailed below.

\begin{figure}[htbp]
    \centering
    \includegraphics[width=\linewidth]{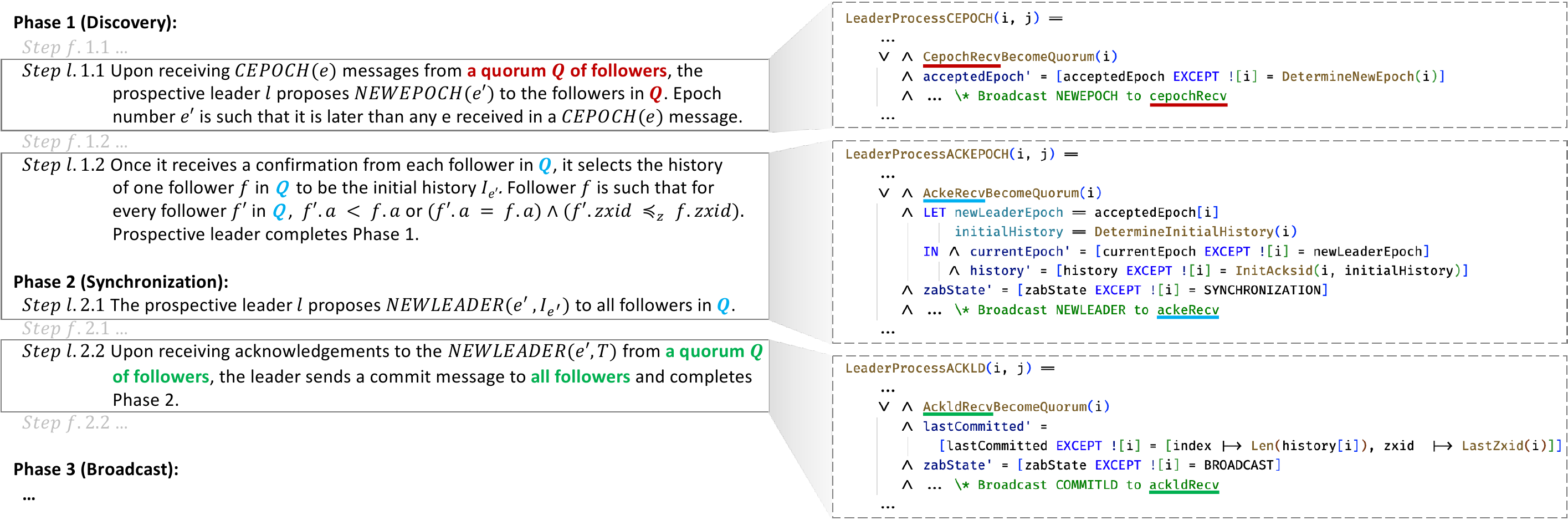}
    \caption{From informal design to formal specification. 
    In the informal design (Left), the set \textsf{Q}, which denotes the quorum of followers, is ambiguous and overloaded. In the \tla\ specification (Right), the set \textsf{Q} is specified with different variables (\textsf{cepochRecv}, \textsf{ackeRecv} and \textsf{ackldRecv}) for different quorums in different situations.}
    \label{F:Prot-Spec}
\end{figure}

First, we need to cope with the ambiguities concerning the abstract mathematical notion of \textit{quorum}.
\zk\ is a leader-based replicated service.
The leader requires certain forms of acknowledgements from a quorum of followers to proceed.
Though the notion of quorum greatly simplifies presentation of the basic rationale of Zab, it introduces subtle ambiguities in the design.
Specifically, the set \textsf{Q}, which denotes the quorum of followers in the Zab pseudocode, is severely overloaded in the informal design. 
It refers to different sets of followers in different cases, as shown in Fig. \ref{F:Prot-Spec}.
In the \tla\ specification, we must use separate variables for different quorums in different situations, e.g., variable \textsf{cepochRecv} for the quorum acknowledging the \textsf{CEPOCH} message and \textsf{ackeRecv} for the quorum acknowledging the \textsf{ACKEPOCH} message.

Second, the design of Zab mainly describes the ``happy case", in which the follower successfully contacts the leader and the leader proceeds with support from a quorum of followers.
In our \tla\ specification, we must precisely describe the unhappy case where the leader has not yet received sufficient acknowledgements from the followers.
We must also explicitly model failures in the execution environment, enabling the TLC model checker to exhaustively exercise the fault-tolerance logic in Zab.
Moreover, Zab is a highly concurrent protocol.
Actions of ZooKeeper processes and environment failures can interleave.
The complex interleavings are not covered in the protocol design,
and we supplement the detailed handling of the interleavings in our \tla\ specification.

Third, the Zab protocol does not implement leader election, but relies on an assumed leader oracle.
In our \tla\ specification, we use a variable called \textsf{leaderOracle} to denote the leader oracle. 
The leader oracle exposes two actions \textsf{UpdateLeader} and \textsf{FollowLeader} to update the leader and to find who is the leader, respectively.

The details of protocol design we supplement in the \tla\ specification are verified by model checking. It is also confirmed by our development of the system specification (see Section \ref{Sec:Syst-Spec}).

\subsubsection{Specification of Correctness Conditions.}
We specify two types of correctness conditions, the \textit{core correctness conditions} and the \textit{invariant properties} (with a focus on safety properties in this work).
The Zab protocol \cite{Junqueira11} prescribes six core correctness conditions, namely \textit{integrity}, \textit{total order}, \textit{agreement}, \textit{local primary order}, \textit{global primary order} and \textit{primary integrity}.
These properties are extracted from the requirement analysis of the \zk\ coordination service.
They constrain the externally observable behavior from the user client's perspective.

Designers usually do not directly guarantee the core correctness conditions when designing a complex protocol like Zab.
Rather, they decompose the core correctness conditions into a collection of invariant properties.
In principle, the invariants maintained in different parts of the protocol can collectively ensure the core correctness conditions.
Model checking against the invariants can also accelerate the detection of bugs in protocol design.
We extract the invariant properties based on our understanding of the Zab protocol.
We are also inspired by the invariants specified for Paxos, Raft and their derivatives \cite{Ongaro14,Paxos-TLA-v1,PaxosTLA2019,Raft-TLA}.


\subsection{Ensuring Quality of Protocol Specification}

Upon completing the development of the protocol specification, we perform model checking to ensure its quality. 
The protocol specification is amenable to automatic exploration by the TLC model checker.
We utilize TLC in two modes: the standard \textit{model-checking mode}, in which TLC traverses the state space in a BFS manner, and the \textit{simulation mode}, where TLC keeps sampling execution paths of a predefined length until it uses up the testing budget.

In this work, we perform model checking on a PC equipped with an Intel I5-9500 quad-core CPU (3.00GHz) and 16GB RAM, running Windows 10 Enterprise. The software used is \tla\ Toolbox v1.7.0. 
We first tame the state explosion problem for the model checking and identify subtle bugs resulting from ambiguities in the informal design. 
Then we conduct further model checking to verify the correctness of the protocol specification.

\subsubsection{Taming State Explosion.} 
Model checking suffers from the state explosion problem, making it impractical to fully check models of arbitrary scales. 
To mitigate this issue, we prune the state space by tuning the enabling conditions of key actions in the protocol specification. 
The basic rationale behind this is to constrain the scheduling of meaningless events. 
For example, too many failure events are meaningless when the leader does not even contact the followers, and such scheduling should be ruled out during the checking process.

Furthermore, we directly limit the state space based on the small-scope hypothesis, which suggests that analyzing small system instances suffices in practice \cite{Maric17}. Specifically, we control the scale of the model by restricting the following configuration parameters: the number of servers, the maximum number of transactions, the maximum number of timeouts, and the maximum number of restarts. The server count is confined to a maximum of 5, as it is sufficient to trigger most invariant violations in most cases according to \zk's bug tracking system \cite{ZK-Issues}. 
Similarly, the number of transactions is limited to a small value, as it already suffices to cause log differences among servers.

It is also worth noting that in the protocol specification, we model failures as \textsf{Timeout} and \textsf{Restart} mainly for state space reduction. 
These two actions can effectively describe the effects of multiple failures in the execution environment.

\subsubsection{Finding Ambiguities in the Informal Design.}
Following the above techniques to mitigate the state explosion,
we perform model checking and find two invariant violations in the preliminary version of our protocol specification.
One is due to the misuse of the quorum of followers.
The other concerns how the leader responds to a recovering follower in between logging a transaction and broadcasting it to the followers.
The paths to the violations help us find the root cause and fix the bugs in the protocol specification.
Our fixes are also in accordance with the system implementation (see Section \ref{Sec:Syst-Spec}), though the implementation contains more details.

\subsubsection{Verifying Correctness.}
After resolving the aforementioned bugs in the protocol specification,  we proceed with model checking to ensure its correctness. 
We adjust the scale of the model using the parameters specified in the model configuration mentioned earlier.
For each configuration, we record the checking mode, the number of explored states, and the checking time cost. 
For the model checking mode, we also record the diameter of the state space graph generated by the TLC model checker.
For the simulation mode, we set the maximum length of the trace to 100.
We restrict the model checking time to 24 hours and the disk space used to 100GB.

\input{./tables/protocol-spec-results.tex}

Table \ref{table:protocol-spec-mck-results} presents statistics regarding the model checking of the protocol specification. 
The explorations shown in the table cover a variety of configurations, with the server count ranging from 2 to 5, and the maximum number of transactions, timeouts, and restarts up to 3. 
Within the time limit of 24 hours, the explorations of all configurations do not exceed the space limit. 
When limiting the model to a relatively small scale, the model-checking mode can exhaustively explore all possible interleavings of actions. 
In contrast, the simulation mode tends to explore deeper states. 
All specified correctness conditions are met without violation during the explorations in models of diverse configurations. 
Based on the results and the small-scope hypothesis \cite{Maric17}, we have achieved a high level of confidence in the correctness of our protocol specification.

Moreover, we further tweak the specification a little and see if the model checker can find the planted errors.
For instance, in one trial, we modified the definition of the constant \textsf{Quorums} in the specification, which originally denotes a set of all majority sets of servers, to include server sets that comprise no less than half of all servers.
In expectation, the modification will lead to invariant violations only when the number of servers is even. 
We executed model checking on the modified specification, and as anticipated, no violations occurred when the server number was 3 or 5. 
However, in the case of 2 or 4 servers, invariant violations emerged, such as two established leaders appearing in the same epoch. 
Such trials illustrate the effectiveness of the specified correctness conditions and further indicate the correctness of the protocol specification. 
More details about the verification of the protocol specification can be found in \cite{ZK-TLA-Spec}.

%% file: tables/protocol-spec-results.tex
\begin{table}
    \caption{Model checking results of the protocol specification.}
    \label{table:protocol-spec-mck-results}
    \centering
   \resizebox{\textwidth}{!}{%
        \centering
        \renewcommand*{\arraystretch}{1}
        \begin{threeparttable}
        \begin{tabular}{c c c c c}
            \toprule
            \textbf{\quad Config\tnote{*} \quad }
            & \textbf{\quad Checking mode \quad  } & \textbf{\quad Diameter \quad  }
            & \textbf{\quad Num of states \quad } & \textbf{\quad Time cost \quad } \\ \hline
            $(2,2,2,0)$        & Model-checking     & $38$  &  $19,980$         &  $00:00:03$  \\ 
            $(2,2,0,2)$        & Model-checking     & $38$  &  $25,959$         &  $00:00:04$  \\ 
            $(2,2,1,1)$        & Model-checking     & $38$  &  $26,865$         &  $00:00:04$  \\ 
            $(2,3,2,2)$        & Model-checking     & $60$  &  $10,370,967$     &  $00:06:58$  \\ 
        
            $(3,2,1,0)$        & Model-checking     & $43$  &  $610,035$        &  $00:00:28$   \\ 
            $(3,2,0,1)$        & Model-checking     & $50$  &  $1,902,139$      &  $00:02:36$   \\
            
            $(3,2,2,0)$        & Model-checking     & $54$  &  $26,126,204$     &  $00:17:07$   \\ 
            $(3,2,0,2)$        & Model-checking     & $68$  &  $245,606,642$    &  $03:41:23$   \\ 
            $(3,2,1,1)$        & Model-checking     & $61$  &  $84,543,312$     &  $01:00:18$   \\ 
            
            $(3,2,2,1)$        & Model-checking     & $50$  &  $1,721,643,089$  &  $>24:00:00$   \\
            $(3,2,1,2)$        & Model-checking     & $46$  &  $1,825,094,679$  &  $>24:00:00$   \\
            
            $(3,3,3,3)$        & Simulation         & $-$   &  $1,194,558,650$  &  $>24:00:00$  \\ 

            $(4,2,1,0)$        & Model-checking     & $64$  &  $21,393,294   $   & $00:23:29$  \\ 
            $(4,2,0,1)$        & Model-checking     & $71$  &  $79,475,010$      & $01:37:31$  \\
            $(4,2,2,0)$        & Model-checking     & $57$  &  $1,599,588,210$   & $>24:00:00$  \\ 
            
            $(5,2,3,3)$        & Simulation         & $- $  &  $1,044,870,264$   & $>24:00:00$   \\
            $(5,3,2,2)$        & Simulation         & $- $  &  $817,181,422  $   & $>24:00:00$  \\ 
            \bottomrule
        \end{tabular}%
        \begin{tablenotes}
        \footnotesize
        \item[*] In the protocol specification, the \textbf{Config} parameters represent the number of servers, the maximum number of transactions, the maximum number of timeouts, and the maximum number of restarts.  
        \end{tablenotes}
        \end{threeparttable}
    }
\end{table}

%% file: sections/4_syst-spec.tex
\section{System Specification} \label{Sec:Syst-Spec}

Given the protocol specification, we further develop the system specification, which serves as the super-doc supplementing detailed system documentation of Zab implementation for the \zk\ developers.
In the following, we first discuss the essentials of a system specification written in \tla.
Then, we present the practice of developing the system specification and the approach to ensuring its quality.

\subsection{Essentials of a Super-doc in \tla}
To develop the system specification, we first need to decide which should and should not be included in the specification.
We fathom out the right level of abstraction from two complementary perspectives, namely what the system developers need from a super-doc and what the \tla\ specification language can express.

As a super-doc, the system specification should reconcile accuracy with simplicity.
When we submitted the preliminary version of the system specification to the \zk\ community \cite{ZK-PR1690}, a basic principle the community suggests is that ``whether or not an action should be retained in the specification depends on whether it is critical to reveal important details of the protocol".
Further suggestions include ``implement the minimum required" and ``keep things simpler".
These suggestions guide us to calibrate the level of abstraction. 
In principle, the system specification should cover every aspect of the \zk\ system.
Meanwhile, low-level details, e.g. the leader-follower heartbeat interactions and internal request queues for blocked threads, can be omitted.
We also inherit the modularity of the system implementation in our specification.

The precision of the specifications written in \tla\ is intended to uncover design flaws.
\tla\ specifications mostly consist of ordinary non-temporal mathematics, e.g., basic set theory, which is less cumbersome than a purely temporal specification. 
A major advantage of \tla\ is that it frees us from complex programming considerations like multi-threading and object encapsulation.
When enjoying the simplicity of \tla, we inevitably need to handle the gap in expressiveness between \tla\ and the programming language (Java in the \zk\ case).
In \zk, the block-and-wakeup style multi-thread programming is heavily used, while in \tla, actions are executed in an asynchronous style. 
We decompose the block-and-wakeup thread logic into two conditional branches in one \tla\ action.
The timing of scheduling the wakeup logic is encoded in the entry conditions of the branches.
Moreover, we combine the wakeup logic of multiple threads in one conditional branch in the \tla\ action. This not only improves specification readability, but also helps mitigate the state explosion.
%

\subsection{Development of the Super-doc}
\begin{figure}[htbp]
    \centering
    \includegraphics[width=\linewidth]{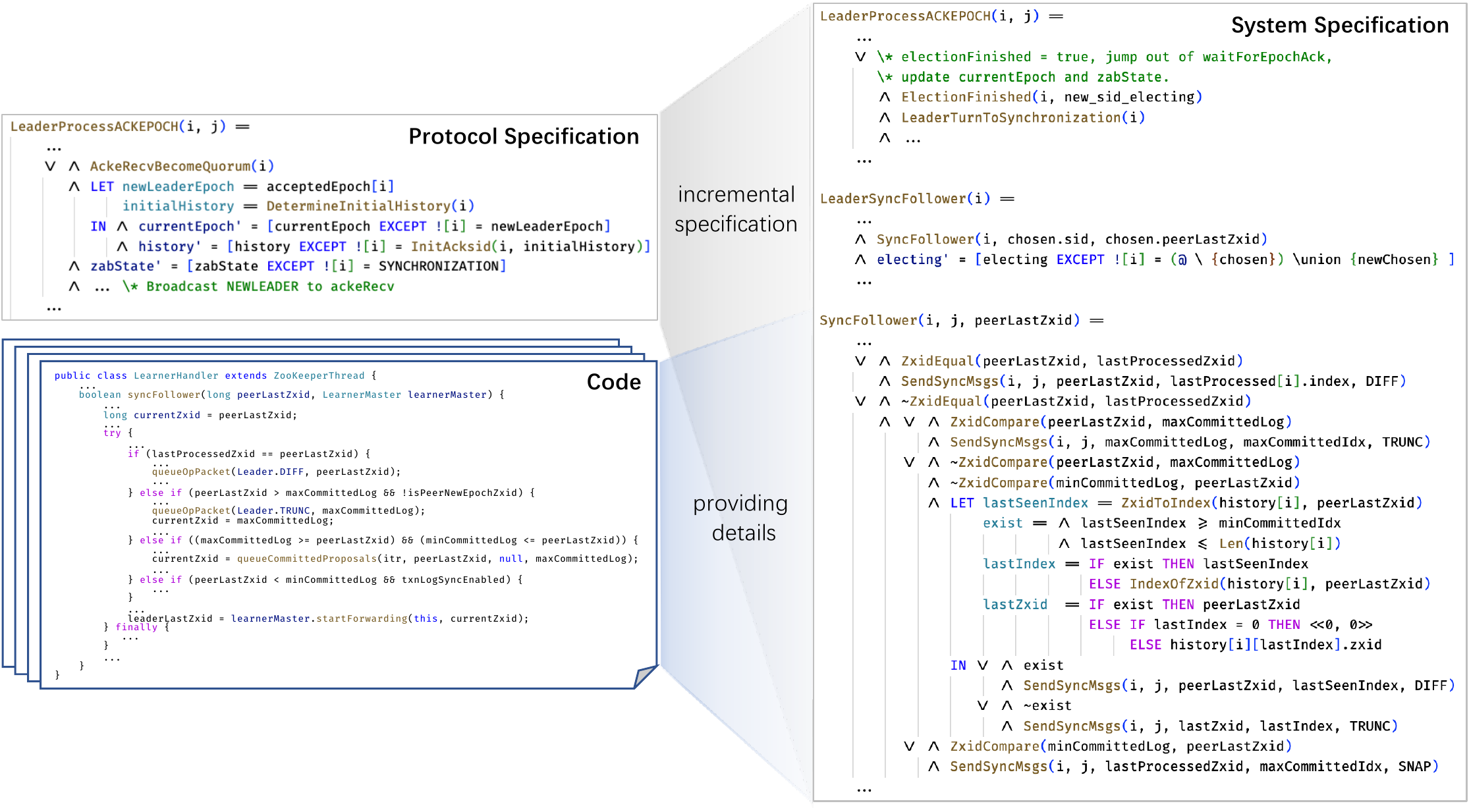}
    \caption{Incremental development of the system specification. 
    The system specification is in principle a refinement of the protocol specification, with supplementary details derived from the source code.}
    \label{F:Syst-Spec}
\end{figure}

The system specification is in principle a refinement of the protocol specification.
Details in the system specification are supplemented based on the source code, as shown in Fig. \ref{F:Syst-Spec}.
\zk\ inherits the basic architecture of Zab, and we discuss each of its modules in turn.

\zk\ implements its own Fast Leader Election (FLE) algorithm \cite{Medeiros12}, which is omitted in Zab.
FLE elects a leader which has the most up-to-date history. 
We also extract the invariant properties FLE should maintain from its design.
Moreover, to conduct ``unit test" on the FLE design, i.e., to model check the FLE module alone, we simulate the interaction between FLE and its outside world, e.g., actions updating the epoch or the transaction ID (\textsf{zxid}).

Compared with Zab, the \textsf{DISCOVERY} module in \zk\ is simplified, since the leader already has the most up-to-date history. 
We reuse most part of this module in the protocol specification, and make several revisions according to the implementation.
Specifically, the follower does not need to send its full history to the leader, and only needs to send the latest \textsf{zxid} it has.
The leader checks the validity of each follower's state rather than updates its history based on all histories received from the followers. 

In Zab, the leader sends its full history to the followers in the \textsf{SYNC} phase.
This is obviously a simplification for better illustration of the design rationale.
In the implementation, the \textsf{SYNC} module is significantly optimized for better performance.
We rewrite the system specification for this module based on the implementation.
Specifically, \textsf{NEWLEADER} acts as a signaling message without carrying concrete history data.
The leader's history will be synchronized to the followers in one of three modes, namely \textsf{DIFF}, \textsf{TRUNC}, and \textsf{SNAP}. 
The leader will select the right mode based on the differences between its history and the follower's latest \textsf{zxid}.
The follower may update its log or snapshot according to the sync mode. 
These supplemented details in the system specification are confirmed by the system implementation.
The \textsf{BROADCAST} module is basically inherited from the protocol specification. 

In order to facilitate conformance checking (see Section \ref{Sec:conformance-checking}), we also refine the failure modeling of the protocol specification. 
Specifically, we model environment failures as node crash/rejoin events and network partition/reconnection events in the system specification. 
These failure events are more fundamental and can generate the failures modeled in the protocol specification. 

Correctness conditions, including core correctness conditions and invariant properties, are mainly inherited from the protocol specification.
Table \ref{table:system-spec-mck-results} presents the model checking results of the system specification under certain configuration parameters.
No violation of correctness conditions is found during the checking.

\input{./tables/system-spec-results.tex}

\subsection{Ensuring Quality of the Super-doc} \label{Sec:conformance-checking}
The quality of the super-doc primarily depends on whether the doc precisely reflects the system implementation, with unimportant details omitted.
Note that model checking can only ensure that the specification satisfies the correctness conditions.
It cannot tell whether the specification precisely reflects the system implementation.

We conduct conformance checking between the system specification and the system implementation to ensure the quality of the specification, as shown in Fig. \ref{F:Test-Spec}.
We first let the TLC model checker execute the system specification and obtain the model-level execution traces.
We extract the event schedule from the model checking trace, and then control the system execution to follow the event schedule.

The controlled system execution is enabled by the Model Checking-driven Explorative Testing (\met) framework \cite{MET}.
We first instrument the \zk\ system, which enables the test execution environment to intercept the communication between the \zk\ nodes, as well as local events of interest, e.g., logging transactions.
The intercepted events are dispatched according to the event schedule extracted from the model checking trace.
In this way, we control the system execution to ``replay" the model checking trace, and check the conformance between these two levels of executions.

Once the conformance checking fails, discrepancies between the specification and the implementation are detected.
The specification developer checks the discrepancies and revises the specification based on the implementation.
After multiple rounds of conformance checking, the system specification obtains sufficient accuracy.
This process is analogous to the regression testing \cite{Bourque14}.

During our practice, we discovered several discrepancies between the specification and the implementation. 
For example, in the initial version of the system specification, it was assumed that whenever the leader processes a write request, the client session has already been successfully established. 
However, client session creation is also considered a transaction in \zk\ and requires confirmation by a quorum of servers. 
This discrepancy was identified during conformance checking, and the specification was subsequently revised to address it.
Further details about the system specification can be found in \cite{ZK-TLA-Spec}.

%% file: tables/system-spec-results.tex
\begin{table}
    \caption{Model checking results of the system specification.}
    \label{table:system-spec-mck-results}
    \centering
    \resizebox{\textwidth}{!}{%
        \centering
        \renewcommand*{\arraystretch}{1}
        \begin{threeparttable}
        \begin{tabular}{c c c c c}
            \toprule
            \textbf{\quad Config\tnote{*} \quad }
            & \textbf{\quad Checking mode \quad  } & \textbf{\quad Diameter \quad  }
            & \textbf{\quad Num of states \quad } & \textbf{\quad Time cost \quad } \\ \hline
            $(3,2,3,3)$         & Model-checking    & $24$  & $3,322,996,126$ & $>24:00:00$ \\ 
            $(5,2,3,3)$         & Model-checking    & $16$  & $693,381,547  $ & $>24:00:00$ \\ 
            $(3,5,5,5)$         & Simulation        & $ -$  & $1,139,420,778$ & $>24:00:00$ \\ 
            $(5,5,5,5)$         & Simulation        & $ -$  & $1,358,120,544$ & $>24:00:00$ \\     
            $(3,5,0,10)$        & Simulation        & $ -$  & $1,463,314,104$ & $>24:00:00$ \\ 
            $(3,5,10,0)$        & Simulation        & $ -$  & $1,211,089,513$ & $>24:00:00$ \\   
            \bottomrule
        \end{tabular}%
        \begin{tablenotes}
        \footnotesize
        \item[*] In the system specification, the \textbf{Config} parameters represent the number of servers, the maximum number of transactions, the maximum number of node crashes, and the maximum number of network partitions. 
        \end{tablenotes}
        \end{threeparttable}
    }
\end{table}

%% file: sections/5_test-spec.tex
\section{Test Specification} \label{Sec:Test-Spec}

\begin{figure}[htbp]
    \centering
    \includegraphics[width=\linewidth]{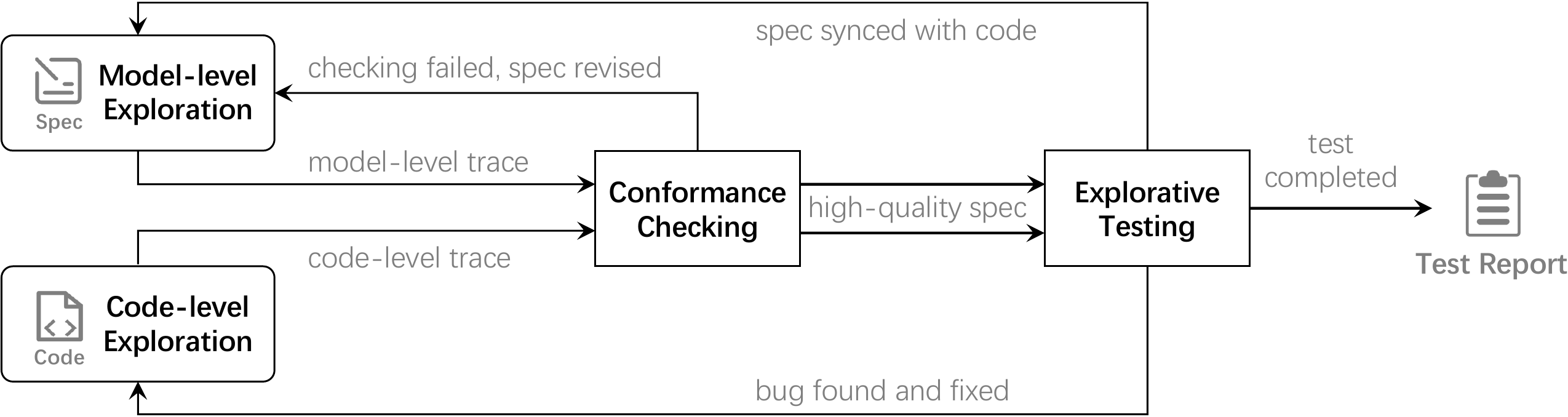}
    \caption{Model checking-driven explorative testing.}
    \label{F:Test-Spec}
\end{figure}

Given the protocol and system specifications, we further develop  the test specification, in order to guide the explorative testing of \zk.
The basic workflow of explorative testing following the \met\ framework is shown in Fig. \ref{F:Test-Spec}.
We detail the three core parts of the framework below.

\subsection{Obtaining the Test Specification}
In the explorative testing, we mainly focus on the recovery logic of \zk\ in its SYNC phase (though \met\ can be applied to each module of \zk).
This module is heavily optimized in practice. 
It is under continual revision and deep bugs can still be found, according to \zk's bug tracking system \cite{ZK-Issues}.

The test specification is in principle a refinement of the system specification toward the source code. 
The test specification first inherits the system specification.
The part of the specification to be refined, which corresponds to the part of the system under test, can be obtained by copy-pasting the \zk\ source code and manually updating the syntax to be valid \tla\ specification \cite{Davis20vldb}.
Due to the inherent gap in the expressiveness, certain details are inevitably omitted or transformed in the test specification, including low-level programming considerations like multi-threading. 
The developer can also intentionally omit certain details if they deem such details irrelevant to the potential deep bugs. 
For example, we do not explicitly model the client and ``pack" the workload generation logic inside the leader.

Due to the state explosion, we cannot model check the test specification of the whole \zk\ system.
In practice, we follow the Interaction-Preserving Abstraction (IPA) framework \cite{Gu22}.
We refine one single module to its test specification, while keeping other modules as abstract as possible, with the interactions between modules preserved.
As we conduct \met\ on the \textsf{SYNC} module, we abstract all other modules.
For example, we combine \textsf{ELECTION} and \textsf{DISCOVERY} into one action, while their interactions with the \textsf{SYNC} module are preserved.

\subsection{Improving the Accuracy of Specification}
The quality of the testing specification is also incrementally improved by multiple rounds of conformance checking (see Section \ref{Sec:Syst-Spec}).
Typically, we find a number of discrepancies.
The developer may need to fine-tune the test specification to better conform to the code.
He may also neglect the discrepancy if he confirms that the discrepancy is due to the inherent differences in expressiveness and is irrelevant to the potential deep bug we try to find.
The conformance checking keeps going like the regression testing until no discrepancy can be found.
This means that the test specification is (sufficiently) accurate to guide the explorative testing.

\subsection{Test Specification in Action}

With the help of our test specification and the \met\ framework, we stably trigger and reproduce several critical deep bugs in \zk\ \cite{ZK-TLA-Spec}.  
Here, we use ZK-2845 \cite{ZK-Issue2845} and ZK-3911 \cite{ZK-Issue3911} as examples to demonstrate the effectiveness and efficiency of this approach. 
These two bugs will result in the inconsistency of accepted logs or the loss of committed logs, which can be particularly severe for a coordination service like \zk. 
However, similar to other deep bugs, they are difficult to uncover and reproduce. 
Triggering these bugs typically requires numerous steps, and the timing of failures is subtle. 
The space of all possible bug-triggering patterns is so vast that it is beyond human reasoning. 
We can only find the bugs by explorative search guided by model checking.

Table \ref{table:test-spec-mck-result} lists the statistics related to the invariant violations of ZK-2845 and ZK-3911.
As indicated, we can obtain the traces of these two bugs within a short time by model checking against the invariants. 
The high efficiency is mainly attributed to the test specification, which abstracts irrelevant details while preserving the necessary information.

\input{./tables/test-spec-results.tex}

Trace analysis reveals that a bug may violate multiple invariants that represent different aspects of the system's requirements. 
For instance, for the two invariants violated by ZK-3911, \textsf{LeaderLogCompleteness} constrains the internal behavior from the developer's perspective,
while \textsf{MonotonicRead} constrains the externally observable behavior from the user client's perspective. 
The quality of the invariants specified in the test specification significantly affects the efficiency of bug detection.
Typically, invariants that constrain internal behaviors can expedite the bug-triggering process compared to the invariants that constrain external behaviors.

The two checking modes of TLC exhibit different capabilities in triggering invariant violations. 
In most cases, the simulation mode is typically faster and more effective in detecting deeper bugs since it tends to explore deeper states.
Conversely, the model-checking mode is better suited for searching for the shortest trace that leads to an invariant violation.

It is worth noting that exposing these bugs through model checking on the system specification can be challenging (see Table \ref{table:system-spec-mck-results}). 
The human knowledge behind the development of the test specification plays a crucial role in pruning the state space and accelerating the bug-triggering process. 
With the flexibility to adjust levels of abstraction in \tla, one can generate the test specification from the system specification at a low cost. 
Besides, TLC enables efficient explorations without additional human effort, and \met\ allows us to replay the traces of invariant violations in the system to validate their authenticity. 
More bugs exposed by our approach are detailed in \cite{ZK-TLA-Spec}.

%% file: tables/test-spec-results.tex
\begin{table}
    \caption{Invariant violations of ZK-2845 and ZK-3911.}
    \label{table:test-spec-mck-result}
    \centering
    \resizebox{\textwidth}{!}{%
        \centering
        \renewcommand*{\arraystretch}{1}
        \begin{threeparttable}
        \begin{tabular}{c|c|c|c|c|c}
            \toprule
            \multirow{2}{*}{\textbf{\ Bug\ }}  & \multirow{2}{*}{\textbf{\ Invariant violation\tnote{*}\ }}  & 
            \multicolumn{2}{c|}{\textbf{Simulation mode}} &    \multicolumn{2}{c}{\textbf{Model-checking mode}}   \\ 
            \cline{3-6}
            ~ & ~  & \textbf{\ Len.\ }  & \textbf{\ Time cost\ }  & \textbf{\ Len.\ }  & \textbf{\ Time cost\ } \\
            \hline
            \multirow{2}{*}{\small{ZK-2845}}  & ProcessConsistency      & 23 & $00:00:02$  & 10 & $00:00:12$      \\ 
            ~                                 & ProposalConsistency     & 20 & $00:00:03$  & 11 & $00:00:18$      \\ 
            \hline
            \multirow{2}{*}{\small{ZK-3911}}  & LeaderLogCompleteness   & 25 & $00:01:29$  & 14 & $00:00:42$      \\ 
            ~                                 & MonotonicRead           & 39 & $00:01:35$  & 18 & $00:13:13$      \\ 
            \bottomrule
        \end{tabular}
        \begin{tablenotes}
        \footnotesize
        \item[*] The column \textbf{Invariant violation} lists the violated invariants of the bugs. The definitions of these invariants can be found in the test specification. The results in the table are obtained using the configuration of 3 servers, 2 transactions in max, 4 node crashes in max, and 4 network partitions in max. 
        \end{tablenotes}
        \end{threeparttable}
    }
\end{table}

%% file: sections/6_related-work.tex
\section{Related Work} \label{Sec:Related-work}

\subsubsection{Specification in \tla.}
\tla\ is widely used for the specification and verification of distributed protocols in both academia and industry. 
Lamport \textit{et al.} utilized \tla\ to specify the original Paxos protocol \cite{PaxosTLA2019}, as well as various Paxos variants, including Disk Paxos \cite{Gafni03}, Fast Paxos \cite{lamport2006fast}, and Byzantine Paxos \cite{lamport2011pluscal}. 
These protocols were also verified to be correct using TLC. 
Diego Ongaro provided a \tla\ specification for the Raft consensus algorithm and further verified its correctness through model checking \cite{Raft-TLA}.
Yin \textit{et al.} employed \tla\ to specify and verify three properties of the Zab protocol \cite{yin2020specification}. 
Moraru \textit{et al.} utilized \tla\ to specify EPaxos when first introducing the protocol \cite{moraru2013there}.

In industry, Amazon Web Services (AWS) extensively employs \tla\ to help solve subtle design problems in critical systems \cite{Newcombe15}. 
Microsoft's cloud service Azure leverages \tla\ to detect deeply-hidden bugs in the logical design and reduce risk in their system \cite{Microsoft-Azure}. 
PolarFS also uses \tla\ to precisely document the design of its ParallelRaft protocol, effectively ensuring the reliability and maintainability of the protocol design and implementation \cite{gu2021raft}. 
WeChat's storage system PaxosStore specifies its consensus algorithm TPaxos in \tla\ and verifies its correctness using TLC to increase confidence in the  design \cite{TPaxos-tla}.

The practices mentioned above utilize \tla\ to specify and verify distributed protocols with the goal of identifying design flaws and increasing confidence in the core protocol design. However, they do not address the code-level implementation and cannot guarantee that the specification accurately reflects the system implementation. 
Discrepancies between the specification and the implementation can result from transcription errors, and model checking is solely responsible for verifying the specification.
Our \tla\ specifications for \zk\ focus on both the protocol design and the system implementation. Based on the source code and the protocol specification, we incrementally develop the system specification that serves as the super-doc for the \zk\ developers. 
Additionally, we conduct conformance checking between the system specification and the system implementation to eliminate discrepancies between them and ensure the quality of the specification.

\subsubsection{Model Checking-Driven Testing on \zk.}
Model checking-driven testing has been extensively employed in distributed systems such as \zk. 
The FATE and DESTINI framework systematically exercises multiple combinations of failures in cloud systems and utilizes declarative testing specifications to support the checking of expected behaviors \cite{gunawi2011fate}. 
This framework has been effectively used to reproduce several bugs in \zk. 
SAMC incorporates semantic information into state-space reduction policies to trigger deep bugs in \zk\ \cite{Lees14}. 
FlyMC introduces state symmetry and event independence to reduce the state-space explosion \cite{lukman2019flymc}. 
PCTCP employs a randomized scheduling algorithm for testing distributed message-passing systems \cite{ozkan2018randomized}, while taPCT integrates partial order reduction techniques into random testing \cite{ozkan2019trace}. 
Both approaches have been utilized to detect bugs in \zk's leader election module. 
Modulo utilizes divergence resync models to systematically explore divergence failure bugs in \zk\ \cite{kimmodulo}. 
The aforementioned works explore \zk\ based on implementation-level model checkers.

In contrast, inspired by the practice of eXtreme Modelling \cite{Davis20vldb} and other test case generation techniques with \tla \cite{Dorminey20,kuprianov2020model}, we leverage the \tla\ specification and the TLC model checker to guide the explorative testing of \zk .
TLC is highly efficient at exploring long traces and uncovering subtle deep bugs that require multiple steps to trigger, making it a powerful tool for test case generation.
Similarly, Mocket uses TLC to guide the testing and reproduces bugs in \zk\ \cite{wang2023model}.
We further reduce the state space by taking advantage of the flexibility of the \tla\ specification, which can be integrated with human knowledge at a low cost.
We develop a test specification that efficiently triggers bugs in \zk, further enhancing the effectiveness of our model checking-driven explorative testing framework.

\bla

%% file: sections/7_concl.tex
\section{Conclusion and Future Work} \label{Sec:Concl}

In this work, we use \tla\ to present precise design of and provide detailed documentation for \zk. We also use model checking to guide explorative testing of \zk.
The formal specifications well complement state-of-the-art testing techniques and further improve the reliability of \zk.

In our future work, we will use \tla\ specifications in more distributed systems, e.g., cloud-native databases and distributed streaming systems.
Enabling techniques, such as taming of state explosion and deterministic simulation of system execution, also need to be strengthened.